\documentclass{aa}

\usepackage{epsfig}
\usepackage{natbib}
\bibpunct{(}{)}{;}{a}{}{,} 

\newbox\grsign \setbox\grsign=\hbox{$>$} \newdimen\grdimen \grdimen=\ht\grsign
\newbox\simlessbox \newbox\simgreatbox \newbox\simpropbox \newbox\wtildebox 
\setbox\simgreatbox=\hbox{\raise.5ex\hbox{$>$}\llap
     {\lower.5ex\hbox{$\sim$}}}\ht1=\grdimen\dp1=0pt
\setbox\simlessbox=\hbox{\raise.5ex\hbox{$<$}\llap
     {\lower.5ex\hbox{$\sim$}}}\ht2=\grdimen\dp2=0pt

\newcommand{\Msun}{\mbox{$M_{\odot}$}}

\newcommand{\be}{\mbox{\begin{equation}}}
\newcommand{\ee}{\mbox{\end{equation}}}

\newcommand{\Mmax}{\mbox{$M_{\rm max}$}}

\newcommand{\Mi}{\mbox{$M_i$}}
\newcommand{\Mp}{\mbox{$M$}}

\newcommand{\tdis}{\mbox{$t_{\rm dis}$}}

\newcommand{\Cref}{\mbox{$m_{\rm ref}$}}

\newcommand{\qev}{\mbox{$q_{\rm ev}$}}

\newcommand{\tfour}{\mbox{$t_4$}}

\newcommand{\nbody}{\mbox{$N$-body}}

\newcommand{\Rc}{\mbox{$R_{\rm CR}$}}

\newcommand{\tds}{\mbox{$t_{\rm ds}$}}
\newcommand{\Vdrift}{\mbox{$V_{\rm drift}$}}
\newcommand{\tdrift}{\mbox{$t_{\rm drift}$}}
\newcommand{\rh}{\mbox{$r_{\rm h}$}}

\newcommand{\tsh}{\mbox{$t_{\rm sp}$}}
\newcommand{\kms}{\mbox{km$\,$s$^{-1}$}}
\newcommand{\pcmyr}{\mbox{pc$\,$Myr$^{-1}$}}
\newcommand{\Vdisk}{\mbox{$V_{\rm disk}$}}
\newcommand{\kmspkpc}{\mbox{km$\,$s$^{-1}$\,${\rm kpc}^{-1}$}}

\newcommand{\tdg}{\mbox{$t_{\rm GMC}$}}

\newcommand{\Msunmyr}{\mbox{$M_{\odot}\,$Myr$^{-1}$}}

\begin{document}

\title{Clusters in the solar neighbourhood: how are they destroyed?}

\author{Henny J.G.L.M. Lamers \inst{1,2} and  Mark Gieles\inst{1}
        }

\institute { 
                 {Astronomical Institute, Utrecht University, 
                 Princetonplein 5, NL-3584CC Utrecht, The Netherlands
                 {\tt  lamers@astro.uu.nl, gieles@astro.uu.nl}}
            \and  {SRON Laboratory for Space Research, Sorbonnelaan 2,
                 NL-3584CC, Utrecht, The Netherlands}
            }

\date{Received date ; accepted date}

\offprints{H. J. G. L. M. Lamers}

\abstract{
We predict the survival time of initially bound star clusters in the solar
neighbourhood taking into account: (1) stellar
evolution, (2) tidal stripping, (3) shocking by spiral arms
and (4) encounters with giant molecular clouds.
We find that the predicted dissolution time is $\tdis= 1.7 (\Mi/10^4\,\Msun)^{0.67}$
Gyr for clusters in the mass range of $10^2 < \Mi < 10^5\,\Msun$.
The resulting predicted shape of the logarithmic age distribution 
agrees very well with the empirical one, derived from 
a complete sample of clusters in the solar neighbourhood within 600 pc.
The required scaling factor implies a star formation rate of $4~10^2$
\Msunmyr\  within 600 pc from the Sun 
or a surface formation rate of   
$3.5~10^{-10}$ \Msun yr$^{-1}$pc$^{-2}$ for stars in bound clusters 
with an initial mass in the range of $10^2$ to $3~10^4$ \Msun.
}

\authorrunning{H.J.G.L.M. Lamers \& M. Gieles}
\titlerunning{The destruction of cluster in the solar neighbourhood} 

\maketitle


\section{Introduction}
\label{sec:1}

 The first empirical determination of the lifetime of clusters in the
solar neighbourhood is by \citet{1958RA......5..507O}, who noticed the
lack of clusters older than a few Gyr in the solar neighbourhood.
Later, \citet{1971A&A....13..309W} derived a mean dissolution time of 0.2
Gyr from the age distribution of clusters.
 Since most of the observed
clusters within about 1 kpc from the sun have a mass in the range of
$10^2$ to a few $10^3$ \Msun\ 
the value derived by Wielen is for clusters in that mass range. Theory
predicts that the dissolution time of clusters depends on their
initial mass in that massive clusters survive longer than low mass
clusters (e.g. \citealt{1958ApJ...127...17S},
\citealt{1985IAUS..113..449W}, \citealt{1990ApJ...351..121C},
\citealt{1997ApJ...474..223G} and references
therein). \citet{2003MNRAS.340..227B} (hereafter BM03) showed from
\nbody\ simulations that the dissolution time of clusters in the tidal
field of the galaxy depends on their initial mass, $M_i$, as $\tdis
\sim M_i^{0.62}$. Independently, this same power-law dependence was
also derived empirically in a study of cluster samples in four
galaxies by \citet{2003MNRAS.338..717B}.

The dissolution time of clusters in the solar neighbourhood was
recently redetermined by \citet{2005A&A...441..117L} (hereafter L05), 
based on a new cluster sample of
\citet{2005A&A...438.1163K}. 
They found a dissolution time of $\tdis =
\tfour (\Mi/10^4 \Msun)^{0.62}$ with $\tfour = 1.3 \pm 0.5$ Gyr
for clusters with  $10^2 < M < 10^4$.
This is a factor 5 shorter than the $\tfour=6.9$ Gyr 
that follows from the
$\tdis(\Mi)$ relation derived
by the \nbody\ simulations of BM03 for clusters more massive than 4500
\Msun\ at a Galactocentric distance of 8.5 kpc.
The simulations of BM03 include a
realistic stellar mass function, stellar evolution, two-body
relaxation, a detailed
treatment of binary evolution and close encounters of stars.
Part of this difference may be due to the fact that in low mass clusters,
with lifetimes shorter than about 1 Gyr, the dynamical evolution is affected 
by the presence of massive stars during most of their lifetime  (see Fig. 5
of BM03). It is doubtfull that this effect alone can fully explain the
difference between the results of L05 and BM03.
In fact, the large discrepancy suggests
that other, probably external,  disruptive effects must play an important role in 
destroying star clusters in the solar neighbourhood.

In this paper we explain the lifetime of clusters in the solar
neighbourhood, by
taking into account the combined effects of stellar evolution, tidal
stripping, encounters with giant molecular clouds (GMCs)
and spiral arm shocks.  We use stellar population models to describe the
stellar evolution and the results of BM03 for tidal stripping. 
For the effects of GMCs and spiral arms we adopt the new
estimates  from the
recent studies by  \citet{gieles06b} (hereafter GPZB06) and
\citet{gieles06a} (hereafter GAPZ06), which are based on \nbody\ simulations.

The structure of the paper is as follows.
In Sect. 2 we discuss the predicted mass loss from star clusters by 
stellar evolution, tidal stripping, encounters with GMCs 
 and spiral arm shocks.
We calculate the mass evolution of clusters due to these
four  effects.
In Sect. 3 we compare our results with the observed age distribution of
clusters in the solar neighbourhood. 
The discussion and conclusions are given in Sect. 4.

\section{The decreasing mass of star clusters}

\subsection{Mass loss by stellar evolution}

The mass loss from clusters due to stellar evolution has been
calculated for cluster evolution models by several groups.  We adopt
the $GALEV$ models for single stellar populations with a Salpeter type
mass function in the range of $0.15 < M/\Msun<85$
\citep{2002A&A...392....1S,2003A&A...401.1063A}. These models are
based on stellar evolution tracks from the Padova group, which include
overshooting,  mass loss due to stellar winds and supernovae.
(\citealt{1994A&AS..106..275B}; \citealt{2000A&AS..141..371G}).  L05
have shown that the fraction of the initial cluster mass that is lost
by stellar evolution, $\qev(t)=\Delta M/\Mi$, can be approximated
accurately by

\begin{equation}
\log \qev(t)= (\log t-a_{\rm ev})^{b_{\rm ev}}+c_{\rm
  ev}~~{\rm for}~~ t>12.5~{\rm Myr} .
\label{eq:qevgot}
\end{equation}
with $t$ in yrs.
For solar metallicity models with a Salpeter mass function $a_{\rm
  ev}=7.00$, $b_{\rm ev}=0.255$ and $c_{\rm ev}=-1.805$.
This function describes the mass loss fraction of the models at
$t>12.5$ Myr with an accuracy of a few percent. The mass loss at 
younger ages is negligible because stars with $M_*>30
\Msun$ hardly contribute to the mass of the cluster.
The mass loss of a cluster by stellar evolution is

\begin{equation}
\left(\frac{dM}{dt}\right)_{\rm evol} = -M(t)\frac{d \qev}{dt}
\label{eq:dmdtevol}
\end{equation}

\subsection{Mass loss by the galactic tidal field}

BM03 have calculated a grid of \nbody\ simulations of clusters in
circular and elliptical orbits in the tidal field of 
a galaxy for
different initial cluster masses, galactocentric distances $R$, and
different cluster density profiles. The stars follow a Kroupa initial mass function
and stellar evolution is taken into account during the evolution. 
\citet{2004ASPC..322..481G} have shown that for all models with
clusters with $\Mi>4500\,\Msun$ of BM03 
the dissolution time can be expressed as a function of the initial
cluster mass as

\begin{equation}
\tdis = \tfour~(\Mi/10^4\,\Msun)^{0.62},
\label{eq:tdis}
\end{equation}
where $\tfour$ is a constant that depends on the tidal field strength
of the galaxy in which the
cluster moves and on the ellipticity of its orbit.
(If the tidal field were the only disruptive process then $\tfour$ would
be the lifetime of a cluster with $\Mi=10^4\,\Msun$.)
The mass loss due to the Galactic tidal field can then be written as

\begin{equation}
\left(\frac{dM}{dt}\right)_{\rm tidal} = \frac{-M(t)}{\tdis} = 
 \frac{-(M/10^4\Msun)^{0.38}}{\tfour / 10^4}\,\Msunmyr
\label{eq:dmdtdis}
\end{equation}
We adopt the value of $\tfour = 6.9\,$Gyr from BM03 for clusters
in circular orbits at $R_0=8.5$ kpc, although this may slightly 
overestimate the disruption time of the lower mass clusters by
about a factor 2 or so (see \S 1).

\subsection{Mass loss by spiral arm shocking}

GAPZ06 studied the dissolution of star clusters by spiral arms by
means of $N$-body simulations. They used and adjusted the analytical
expression of \citet{1972ApJ...176L..51O} for the dissolution time of
star clusters due to disk shock, to derive an expression for the
dissolution time of star clusters by spiral arms ($\tsh$). 
Mass loss by spiral arm shocks will occur just at the moment the cluster
crosses the spiral arm.
Assuming
that spiral arms move with a constant pattern speed ($\Omega_{\rm p}$)
and that the matter in the disk has a constant circular
velocity ($V_{\rm disk}$), the relative velocity between the two
($\Vdrift$) depends on the location in the galaxy ($R$). Density waves
that pass with a low velocity have a large effect on the star clusters
(e.g. \citealt{1972ApJ...176L..51O}). Therefore, the disruptive
effect of spiral arm shocks is most important close to the corotation
radius ($R_{\rm CR}$), i.e. the point where the disk and the spiral
arms have the same rotational velocity. We adopt the ``average''
spiral arm model of GAPZ06, 
which is based on the study 
of
\citet{1989ApJ...343..602E} of the spiral galaxies M81 and M100,  to
derive the density contrast of the spiral arm.
\citet{2005ApJ...629..825D} found $\Omega_{\rm p} =25.9\,\kmspkpc$
for the spiral arms in the Galaxy, from a study of the nearby star
clusters, and a corotation radius ($\Rc$) almost coinciding with the
solar radius $\Rc/R_0=1.06 \pm 0.08$. Based on the adopted values of $R_0 =
8.5\,$kpc, $\Vdisk=220\,\kms$ and the assumption that our Galaxy has 4
spiral arms \citep{2005AJ....130..569V}, GAPZ06 used
$\Vdrift=12.5\,\kms=12.7\,\pcmyr$ and $\tdrift= 1.05\,$Gyr.
Taking into account the ratio $f=(\Delta M/M)/\Delta E/E)=0.3$ between
the energy gain and the mass loss, predicted by GAPZ06 we find for
the solar neighbourhood that
\begin{eqnarray}
\tsh & = & 20\,\left(\frac{M}{10^4\,\Msun}\right)\left(\frac{3.75\,\mbox{pc}}{\rh}\right)^3\,{\rm Gyr}
\label{eq:tdisarm1} \nonumber \\
        & = & 20\,\left(M/10^4\,\Msun\right)^{1-3 \lambda}\,{\rm Gyr},
\label{eq:tdisarm2}
\end{eqnarray}
where we have substituted the observed mass-radius relation of
clusters in nearby spiral galaxies of \citet{2004A&A...416..537L}:
$\rh = 3.75\ (M / 10^4 \Msun)^{\lambda}$, with $\lambda=0.10\pm0.03$.
The mass loss of clusters due to spiral arm shocks is then

\begin{equation}
\left(\frac{dM}{dt}\right)_{\rm sp} =  \frac{-M(t)}{\tds} 
 = -0.5 \left(\frac{M(t)}{10^4\,\Msun}\right)^{3 \lambda}  ~ \Msunmyr.
\label{eq:dmdtarm1}
\end{equation}
Notice that for $\lambda = 0.1$ the mass loss due to shocking by
spiral arms has almost the same mass dependence, i.e. $\propto
M^{0.3}$, as the mass loss by the tidal field (viz. $M^{0.38}$).  
We will use the mass loss relation
Eq.~\ref{eq:dmdtarm1} as a statistical mean.

\subsection{Mass loss by giant molecular cloud encounters}

GPZB06 studied the encounters between GMCs and clusters with $N$-body
simulations. They derived an expression for the energy gain and the
resulting mass loss for the full range of encounter distances, from
head-on to distant encounters.  Adopting 
a mean GMC density in the galactic plane near the sun of
$\rho_n=0.03 \Msun$pc$^{-3}$, a surface density of GMCs
$\Sigma_n=170 \Msun$ pc$^{-2}$ 
\citep{1987ApJ...319..730S} and a mean
velocity dispersion of clusters and GMCs of $\sigma_v \simeq 10~\kms$, they derived a
dissolution time ($\tdg$) for clusters by GMC encounters in the solar
neighbourhood of
  
\begin{equation}
\tdg =  2.0\,\left(\frac{M}{10^4\,\Msun}\right)\left(\frac{3.75 \mbox{ pc}}{\rh}\right)^3{\mbox{Gyr}}.
\label{eq:tdisgmc1}
\end{equation}
If we assume the same mass-radius relation as before we find
that the mass loss rate due to encounters with GMCs is 

\begin{equation}
\left(\frac{dM}{dt}\right)_{\rm GMC} = \frac{-M(t)}{\tdg} =  
-5.0\,\left(\frac{M(t)}{10^4\,\Msun}\right)^{3\lambda} \Msunmyr.
\label{eq:dmdtgmc1}
\end{equation}
Notice that the mass dependence is the same as for dissolution by
spiral arm shocking, but that the effect is ten times stronger.


\subsection{The predicted mass evolution of clusters in the solar
neighbourhood}
\label{subsec:masslosspredictions}

\begin{figure}
\centerline{\psfig{figure=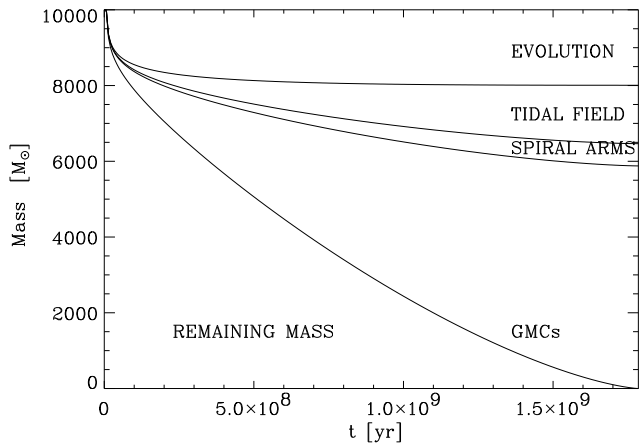, width=9.0cm}}
\caption{The mass  evolution of a cluster with an initial mass of
  $10^4\,\Msun$ in the solar neighbourhood.
The mass loss due to the four separate effects
is indicated. Encounters with GMCs
are the dominant dissolution effect in the solar neighbourhood. }
 \label{fig:1}
\end{figure}

The  decrease of mass due to the combined effects of stellar evolution,
tidal stripping, spiral arm shocks and GMC encounters can then be described as

\begin{equation}
\frac{d\Mp}{dt}= \left(\frac{d\Mp}{dt}\right)_{\rm ev} +
                 \left(\frac{d\Mp}{dt}\right)_{\rm tidal}+
                 \left(\frac{d\Mp}{dt}\right)_{\rm sp} +
                 \left(\frac{d\Mp}{dt}\right)_{\rm GMC},
\label{eq:dmpdt}
\end{equation}
with the terms given by Eqs. \ref{eq:dmdtevol}, \ref{eq:dmdtdis},
\ref{eq:dmdtarm1} and \ref{eq:dmdtgmc1}. We have solved this equation
numerically for clusters of different masses. The results are shown in
Fig. \ref{fig:1} for a cluster with an initial mass of $10^4$ 
\Msun. The figure shows the total mass loss as well as the mass
lost by each mechanism independently. Encounters with GMCs are the
dominant dissolution effect in the solar neighbourhood, 
contributing about as much as the three other effects
combined. 

\begin{figure}
\centerline{\psfig{figure=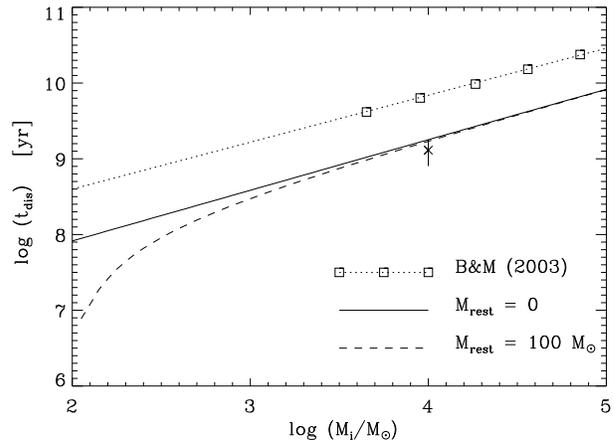, width=9.0cm}}
\caption[]{The predicted dissolution times of clusters in the solar
neighbourhood due to the combined effects of stellar evolution, 
tidal field, spiral arm
shocks and encounters with GMCs, as a function of the initial mass.
 Full line: total dissolution time. Dashed line: time when the
 remaining mass is 100 \Msun. Squares and dotted line: dissolution time due to
 stellar evolution and the Galactic tidal field only, predicted by
 BM03. 
 Cross with error bar: the value of $t_4$ empirically derived by L05.}
\label{fig:2}
\end{figure}

Figure \ref{fig:2} shows the ages of clusters
 when their remaining
mass is 0 and 100 \Msun\
as a function of the initial mass. 
The almost linear part from log(\Mi/\Msun) = 3.5 to 5
has a slope of about 0.67. 
The figure also shows the dissolution times by the Galactic 
tidal field, predicted by
BM03 for clusters with an initial concentration factor $W_0=5$
in a circular orbit at $R_0=8.5$ kpc. Our predicted timescales are about
 a factor 5 smaller, which agrees with the empirically determined dissolution time (L05).

\section{Comparison with observed age distribution of clusters
in the solar neighbourhood}

Given the initial mass distribution of the clusters, their formation
rate, CFR$(t)$, and the time it takes for a dissolving cluster to fade
below the detection limit, we can predict
the distribution of observable clusters as a function of age
or mass.
L05 have derived an expression for the general 
case of a cluster sample that is set by a magnitude limit. Here we are
interested in the prediction for a cluster sample that is complete
down to a mass limit of 100 \Msun, because this is the mass limit of
the unbiased sample of clusters within 600 pc 
of \citet{2005A&A...438.1163K} (see L05).

For a constant CFR and a power law cluster IMF with
a slope of $-\alpha=-2$ \citep{2003ARA&A..41...57L} the number of clusters with
$M>100~\Msun$ as a function of age is

\begin{equation}
N_{>100}(t) = C ~ ( M_{\rm lim}(t)^{-1} - M_{\rm max}^{-1} ),
\label{eq:Nt}
\end{equation}
where $M_{\rm lim}(t)$ is the {\it initial} mass of clusters that
reach $M(t)=100\,\Msun$ at age $t$.  (Clusters of age $t$ with a smaller initial
mass have $M<100\,\Msun$ by now.) $\Mmax$ is the maximum {\it initial}
mass of the clusters that are formed. The constant $C$ is related to the star
formation rate (SFR) in bound clusters as 
${\rm SFR}=C \ln (M_{\rm max}/M_{\rm lim})$ for $-\alpha=-2$.

\begin{figure}
\centerline{\psfig{figure=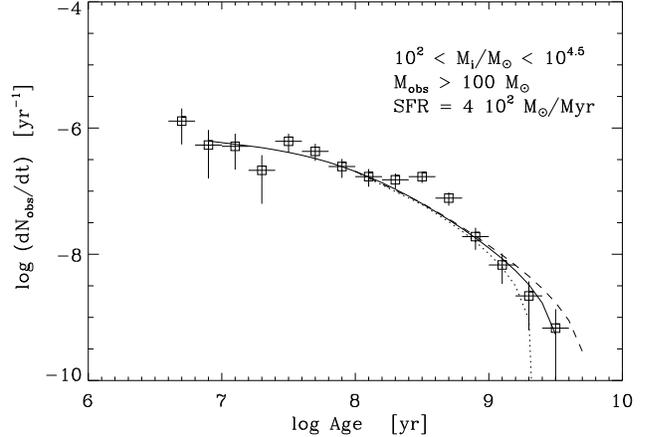,width=9.0cm}}
\caption[]{The observed age distribution of an unbiased sample
of  clusters with $M>100\, \Msun$ in the solar neighbourhood within 600 pc (Karchenko
et al. 2005; L05) in units of nr~yr$^{-1}$ is given by squares with the
Poisson error bars. The full line shows the predicted distribution
for a cluster sample with a maximum mass of $M_{\rm max}=3~10^4~\Msun$
and a SFR of $4~10^2\,\Msunmyr$. 
The dotted line is for $M_{\rm max}=1.5~10^4~\Msun$ and the
  dashed line is for $M_{\rm max}=6~10^4~\Msun$.
}
\label{fig:3}
\end{figure}

Figure \ref{fig:3} shows a comparison between the observed age distribution of clusters
with $M>100\,\Msun$ within 600 pc (from L05)  
with the predicted distribution for
$\Mmax=3~10^4 \Msun$. This value of \Mmax\ is adopted because the
observed distribution shows a steep drop at $\log~t \simeq 9.5$  (with
only one cluster in the last bin) and
Fig. \ref{fig:2} shows that this corresponds to $\Mi = 3~10^4\,\Msun$.
The predicted relation  for twice higher or lower values of $\Mmax$ 
agree  worse with the observed relation. However, see discussion in Sect. 4. 
We have also calculated the expected age distribution in case
  there was no mass-radius relation for the clusters, i.e. for
  $\lambda=0$. The downward slope of the resulting distribution (not
  shown here) is significantly less steep than the one for
  $\lambda=0.1$ and does not fit the observed distribution.

The flattening of the predicted distribution at the low age end is due to the
fact that clusters with an initial mass in the range of
about 100 to 300 \Msun\ quickly reach 100
\Msun\ (see Fig. \ref{fig:2}). The bump in the observed distribution
around $\log (t)\simeq 8.5$ is due to a local starburst (see L05 and 
\citet{2006A&A...445..545P}).
Notice the good agreement in the shapes of the predicted and observed
distributions!

The vertical shift that is applied to the predicted curve
to match the observed one gives a value of $C=10^{-4.15}$ in
Eq. \ref{eq:Nt},
which corresponds to a SFR of $4~10^2\,\Msunmyr$ for bound clusters in the range of 
$10^2 < \Mi/\Msun < 3~10^4$ within 600 pc from the sun.

\section{Discussion and summary}

We studied the dissolution of star clusters in the solar neighbourhood
due to four effects: stellar evolution, tidal stripping, spiral arm
shocks and encounters with GMCs. For this study we adopted the 
descriptions of GAPZ06 and GPZB06 for the dissolution of star
clusters by spiral arms and encounters with GMCs.
We found that the last effect plays a  dominant role in the solar
neighbourhood.

The cluster dissolution time due to spiral arms and GMCs depends
on the density of the clusters, i.e. on $M/\rh^3$. This implies that
the dissolution time is $\tdis \sim M^{\gamma}$ with
$\gamma={1-3\lambda}$ if the radius of a cluster depends on its mass
as $\rh \sim M^{\lambda}$. We adopted $\lambda \simeq 0.1$ as found by
\citet{2004A&A...416..537L} for clusters in spiral galaxies and so
$\gamma \simeq 0.7$ for dissolution by both spiral arms and GMCs. This
value is very similar to $\gamma=0.62$ predicted for dissolution by
the tidal field only (BM03) and empirically derived for cluster
samples in four galaxies by \citet{2003MNRAS.338..717B}.  
If there was
no mass radius dependence for clusters in the solar neighbourhood,
 e.g. $\lambda=0$, then the predicted age distribution would have
a shallower slope than shown in Fig. \ref{fig:3}, since 
more old (massive) clusters would have survived.

 Our calculated dissolution times of clusters in the solar
neighbourhood are about a factor five smaller than predicted by BM03
for clusters in the tidal field of our Galaxy, with stellar evolution,
binaries and two-body relaxation taken into account.
This is reminiscent of
the short dissolution time of clusters in the central region of the
interacting galaxy M51, where the empirical dissolution time is 
even ten times shorter than can be explained by stellar evolution and
tidal fields \citep{2005A&A...441..949G}. GMCs severely limit the
lifetime of clusters in that galaxy also (see the discussion in
GPZB06).

The steep drop in the observed age distribution at $t \simeq 2$ Gyr
can be explained by an upper mass limit for the initial cluster
mass in the solar neighbourhood of about $3~10^4$ \Msun. However, this
value is uncertain because it depends crucially on the completeness of
the used sample at ages above 1 Gyr. (The sample contains only six
clusters older than 1 Gyr.)  The mass versus age distribution of
our adopted sample, shown in Fig. 8 of L05, suggests that the lower
mass limit of the observed clusters increases steeply for clusters
older than 1 Gyr. Since the predicted value of $dN/dt$ at any age
depends on $M_{\rm lim}$ and $M_{\rm max}$ as given in
Eq. \ref{eq:Nt}, and $M_{\rm lim}$ increases when the lower mass limit
increases, an increase in this limit implies an increase in 
$M_{\rm max}$ derived from the observed age distribution. 
Based on this argument
and the small number of clusters older than about 1 Gyr
in the observed sample, we  conclude that the derived value of 
$M_{\rm max}= 3~10^4\,\Msun$ should be considered as a lower limit of
the real maximum initial mass.

The vertical shift applied to the  predicted age distributions to
match the observed one indicates a star formation rate of $4~10^2$
\Msunmyr\ in bound clusters of $\Mi>100\,\Msun$ within a distance of
600 pc, corresponding to a surface formation rate of $3.5~10^{-10}$
\Msun yr$^{-1}$pc$^{-2}$. This is a factor 2 to 3 smaller than the SFR
derived from the study of embedded stars by \citet{2003ARA&A..41...57L}
because many of the stars are born in unbound clusters that dissolve
within 10 Myr.

The very good agreement between the predicted and observed age
distribution of clusters shows that dissolution of clusters in the
solar neighbourhood is dominated by encounters with GMCs, as was
already suggested by \citet{1958RA......5..507O}.
 In fact, the good agreement may be slightly fortuitous because we have 
underestimated the dissolution by two-body relaxation (see
\S 1) and slightly overestimated the dissolution by encounters
with GMCs, because we adopted the midplane density of GMCs whereas clusters
may spend a faction of their lifetime above or under the galactic
disk.
Both effects are expected to be smaller than a factor two and
may partially cancel out.


\section*{Acknowledgments}

This work is supported by a grant from the Netherlands Research
School for Astronomy (NOVA).


\bibliographystyle{aa}

\end{document}